\documentclass[a4paper]{jpconf}
\usepackage{graphicx}
\begin{document}

%\usepackage{amsmath,srcltx}
%\usepackage{amsfonts,amssymb,latexsym}
%\usepackage{mathrsfs}%
%%\usepackage{amscd}
%\usepackage{eucal}
%\usepackage{amsthm}
%\usepackage{epsfig}  %% TENTATIVA - TEM CLASH WITH PACKAGE GRAPHICS
%% MAS PARECE FUNCIONAR
%\usepackage[dvips]{graphics,color}
%\usepackage{axodraw}
%\hfuzz5pt % Don't bother to report overfull boxes < 5pt
%%%%%%%%%%%%%%%%%%%%%%%%%%%%%%%%%%%%%%%%%%%%%%%%%%%%%%%%
%\input portug

\newcommand{\pic}{$\spadesuit$}
\newcommand{\para}{$\bullet$}

\def\aa{$^{{\rm \underline{a}}}$}
\def\aas{$^{{\rm \underline{as}}}$}
\def\oo{$^{{\rm \underline{o}}}$}

\def\ii{\'\i}
\def\ao{\~ao}
\def\oes{\~oes}
\def\oe{\~oe}
\def\cao{\c c\~ao}
\def\coes{\c c\~oes}
\def\coe{\c c\~oe}
\def\an{\~a}
\def\ae{\~ae}
\def\AO{{\~AO}}
\def\OE{{\~OE}}
\def\OES{{\~OES}}
\def\AE{{\~AE}}
\def\CAO{{{\c C}\~AO}}
\def\COE{{{\c C}\~OE}}
\def\COES{{{\c C}\~OES}}

%%%%%%%%%%%%%%%%%%%%%%%%%%%%%%%
%  fichier OL2000.STY pour LATEX (SEM FORMATAGEM)
%%%%%%%%%%%%%%%%%%%%%%%%%%%%%%%%%%%
%\renewcommand{\r}{\rho} %!!!!!!!!!!!!!!
%%%%%%%%%%%%%%%%%%%%%%%%%%%%%%%%%%%%%%%%%%%%%%%%%%
%ACENTOS
\def\aa{$^{{\rm \underline{a}}}$}
\def\aas{$^{{\rm \underline{as}}}$}
\def\oo{$^{{\rm \underline{o}}}$}

\def\ii{\'\i}
\def\ao{\~ao}
\def\oes{\~oes}
\def\oe{\~oe}
\def\cao{\c c\~ao}
\def\coes{\c c\~oes}
\def\coe{\c c\~oe}
\def\an{\~a}
\def\ae{\~ae}
\def\Ii{{\'\I}}
\def\AO{{\~AO}}
\def\OE{{\~OE}}
\def\OES{{\~OES}}
\def\AE{{\~AE}}
\def\CAO{{{\c C}\~AO}}
\def\COE{{{\c C}\~OE}}
\def\COES{{{\c C}\~OES}}
%**************************************************************
\def\ftoday{{\sl {Le \number\day \space\ifcase\month 
\or janvier\or f\'evrier\or mars\or avril\or mai
\or juin\or juillet\or ao\^ut\or septembre\or octobre
\or novembre \or d\'ecembre\fi\space \number\year}}}    
%*****************************************************************
\def\ptoday{{\sl {\number\day \space de\space \ifcase\month 
\or janeiro\or fevereiro\or mar{\c c}o\or abril\or maio
\or junho\or julho\or agosto\or setembro\or outubro
\or novembro \or dezembro\fi\space de\space \number\year}}}    
%*****************************************************************
\def\gtoday{{\sl {Den \number\day. \ifcase\month 
\or Januar\or Februar\or M\"arz\or April\or Mai
\or Juni\or Juli\or August\or September\or Oktober
\or November \or Dezember\fi\space \number\year}}}    
%*****************************************************************
\def\today{{\sl {\ifcase\month
\or January\or February\or March\or April\or May
\or June\or July\or August\or September\or October
\or November \or December\fi \space\number\day,\space 
                                            \number\year}}}
%*****************************************************************
%JOURNAUX
\newcommand{\journal}[4]{{\em #1~}#2\,(#3)\,#4}
\newcommand{\aihp}{\journal {Ann. Inst. Henri Poincar\'e}}
\newcommand{\jhep}{\journal {J. High Energy Phys.}}
\newcommand{\hpa}{\journal {Helv. Phys. Acta}}
\newcommand{\sjpn}{\journal {Sov. J. Part. Nucl.}}
\newcommand{\ijmp}{\journal {Int. J. Mod. Phys.}}
\newcommand{\ijtp}{\journal {Int. J. Theor. Phys.}}
\newcommand{\physu}{\journal {Physica (Utrecht)}}
\newcommand{\pr}{\journal {Phys. Rev.}}
\newcommand{\jetpl}{\journal {JETP Lett.}}
\newcommand{\prl}{\journal {Phys. Rev. Lett.}}
\newcommand{\jmp}{\journal {J. Math. Phys.}}
\newcommand{\jp}{\journal {J. Phys.}}
\newcommand{\rmp}{\journal {Rev. Mod. Phys.}}
\newcommand{\cmp}{\journal {Commun. Math. Phys.}}
\newcommand{\ctp}{\journal {Commun. Theor. Phys.}}
\newcommand{\cqg}{\journal {Class. Quantum Grav.}}
\newcommand{\zp}{\journal {Z. Phys.}}
\newcommand{\np}{\journal {Nucl. Phys.}}
\newcommand{\pl}{\journal {Phys. Lett.}}
\newcommand{\mpl}{\journal {Mod. Phys. Lett.}}
\newcommand{\prep}{\journal {Phys. Rep.}}
\newcommand{\ptp}{\journal {Progr. Theor. Phys.}}
\newcommand{\nc}{\journal {Nuovo Cimento}}
\newcommand{\app}{\journal {Acta Phys. Pol.}}
\newcommand{\apj}{\journal {Astrophys. Jour.}}
\newcommand{\apjl}{\journal {Astrophys. Jour. Lett.}}
\newcommand{\annp}{\journal {Ann. Phys. (N.Y.)}}
\newcommand{\jgp}{\journal {J. Geom. Phys.}}
\newcommand{\Nature}{{\em Nature}}
\newcommand{\PRD}{{\em Phys. Rev. D}}
\newcommand{\MNRAS}{{\em M. N. R. A. S.}}
%**********************************************************************
%LETTRES GRECQUES
\renewcommand{\a}{\alpha}
\renewcommand{\b}{\beta}
\newcommand{\g}{\gamma}           \newcommand{\GA}{\Gamma}
\renewcommand{\d}{\delta}         \newcommand{\D}{\Delta}
\newcommand{\ka}{\kappa}
\newcommand{\la}{\lambda}        \newcommand{\LA}{\Lambda}
\newcommand{\m}{\mu}
\newcommand{\n}{\nu}
\newcommand{\om}{\omega}         \newcommand{\OM}{\Omega}
\newcommand{\p}{\psi}             %\newcommand{\PS}{\Psi} 
\newcommand{\s}{\sigma}           \renewcommand{\S}{\Sigma}
\newcommand{\f}{{\phi}}           \newcommand{\F}{{\Phi}}
\newcommand{\vf}{{\varphi}}
\newcommand{\y}{{\upsilon}}       \newcommand{\Y}{{\Upsilon}}
\newcommand{\z}{\zeta} 
%************************************************************************
%LETTRES SCRIPTES
\renewcommand{\AA}{{\cal A}}
\newcommand{\BB}{{\cal B}}
\newcommand{\CC}{{\cal C}}
\newcommand{\DD}{{\cal D}}
\newcommand{\EE}{{\cal E}}
\newcommand{\FF}{{\cal F}}
\newcommand{\GG}{{\cal G}}
\newcommand{\HH}{{\cal H}}
\newcommand{\II}{{\cal I}}
\newcommand{\JJ}{{\cal J}}
\newcommand{\KK}{{\cal K}}
\newcommand{\LL}{{\cal L}}
\newcommand{\MM}{{\cal M}}
\newcommand{\NN}{{\cal N}}
\newcommand{\OO}{{\cal O}}
\newcommand{\PP}{{\cal P}}
\newcommand{\QQ}{{\cal Q}}
\newcommand{\SSS}{{\cal S}}
\newcommand{\RR}{{\cal R}}
\newcommand{\TT}{{\cal T}}
\newcommand{\UU}{{\cal U}}
\newcommand{\VV}{{\cal V}}
\newcommand{\WW}{{\cal W}}
\newcommand{\XX}{{\cal X}}
\newcommand{\YY}{{\cal Y}}
\newcommand{\ZZ}{{\cal Z}}
%***********************************************************************
%SIGNES SPECIAUX
%\newcommand{\esp}{\\[3mm]}
\newcommand{\esp}{\\[2mm]}
\newcommand{\ES}{\\[6mm]}
\newcommand{\dis}[1]{\displaystyle{#1}}
\newcommand{\na}{\nabla}
\newcommand{\xint}{\dint d^4x\;}
\newcommand{\sla}{\raise.15ex\hbox{$/$}\kern -.57em} 
\newcommand{\Sla}{\raise.15ex\hbox{$/$}\kern -.70em}
\def\h{\hbar}
\def\Lp{\displaystyle{\biggl(}}
\def\Rp{\displaystyle{\biggr)}}
\def\LP{\displaystyle{\Biggl(}}
\def\RP{\displaystyle{\Biggr)}}
\newcommand{\lp}{\left(}\newcommand{\rp}{\right)}
\newcommand{\lc}{\left[}\newcommand{\rc}{\right]}
\newcommand{\lac}{\left\{}\newcommand{\rac}{\right\}}
\newcommand{\identity}{\bf 1\hspace{-0.4em}1}
\newcommand{\complex}{{\kern .1em {\raise .47ex
\hbox {$\scriptscriptstyle |$}}
    \kern -.4em {\rm C}}}
\newcommand{\real}{{{\rm I} \kern -.19em {\rm R}}}
\newcommand{\rational}{{\kern .1em {\raise .47ex
\hbox{$\scripscriptstyle |$}}
    \kern -.35em {\rm Q}}}
\renewcommand{\natural}{{\vrule height 1.6ex width
.05em depth 0ex \kern -.35em {\rm N}}}
\newcommand{\tint}{\int d^3 \! x \, }
\newcommand{\intm}{\int_\MM}
\newcommand{\trace}{{\rm {Tr} \,}}
\newcommand{\half}{\frac{1}{2}}
\newcommand{\pa}{\partial}
\newcommand{\pad}[2]{{\frac{\partial #1}{\partial #2}}}
\newcommand{\fud}[2]{{\frac{\delta #1}{\delta #2}}}
\newcommand{\dpad}[2]{{\displaystyle{\frac{\partial #1}{\partial #2}}}}
\newcommand{\dfud}[2]{{\displaystyle{\frac{\delta #1}{\delta #2}}}}
\newcommand{\dfrac}[2]{{\displaystyle{\frac{#1}{#2}}}}
\newcommand{\dsum}[2]{\displaystyle{\sum_{#1}^{#2}}}   
\newcommand{\dint}{\displaystyle{\int}}
\newcommand{\eg}{{\em e.g.,\ }}
\newcommand{\Eg}{{\em E.g.,\ }}
\newcommand{\ie}{{{\em i.e.},\ }}
\newcommand{\Ie}{{\em I.e.,\ }}
\newcommand{\nb}{\noindent{\bf N.B.}\ }
\newcommand{\etc}{{\em etc.\ }}
\newcommand{\via}{{\em via\ }}
\newcommand{\cf}{{\em cf.\ }}
\newcommand{\twiddle}{\lower.9ex\rlap{$\kern -.1em\scriptstyle\sim$}}
\newcommand{\grad}{\nabla}
\newcommand{\bra}[1]{\left\langle {#1}\right|}
\newcommand{\ket}[1]{\left| {#1}\right\rangle}
% INPUT FILE ol99.sty. PLEASE DO NOT MODIFY IT! (Olivier Piguet)

\newcommand{\vev}[1]{\left\langle {#1}\right\rangle}
%***************************************************************************
%EQUATIONS
\newcommand{\equ}[1]{(\ref{#1})}
\newcommand{\eq}{\begin{equation}}
\newcommand{\eqn}[1]{\label{#1}\end{equation}}
\newcommand{\eea}{\end{eqnarray}}
\newcommand{\eqa}{\begin{eqnarray}}
\newcommand{\eqan}[1]{\label{#1}\end{eqnarray}}
\newcommand{\ba}{\begin{array}}
\newcommand{\ea}{\end{array}}
\newcommand{\eqac}{\begin{equation}\begin{array}{rcl}}
\newcommand{\eqacn}[1]{\end{array}\label{#1}\end{equation}}
\newcommand{\qq}{&\qquad &}
\renewcommand{\=}{&=&} %seems not to work in footnotes!!!           
%---------------  FIN  --------------%

%%%%%%%%%%%%%%%%%%%%%%%%%%%%%%%%%%%%%%%%%%
%%%%%%%%%%%%%%%%%%%%%%%%%%%%%%%%%%%%%%%%%%
%%%%%%%%%%%%%%%%%%%%%%%%%%%%%%%%%%%%%%%%%%
\def\Lac{\displaystyle{\Biggl\{}}
\def\Rac{\displaystyle{\Biggr\}}}
\newcommand{\li}{\\ \hspace{5mm}}
\newcommand{\intx}{{\dint} d^4x\,}
\newcommand{\bz}{\begin{enumerate}}
\newcommand{\ez}{\end{enumerate}}
\newcommand{\doint}{\displaystyle{\oint}}

%\newcommand{\parag}{\\$\ $\\}
%%%%%%%%%%%%%%%%%%%%%%%%%%%%%% NEW COLOR DEFINITIONS %%%%%%%%%%%%%%%%%%%%%%%%%%%%%%%
%\definecolor{azulinho}{rgb}{0.25,0,0.9}
%\definecolor{verdelinho}{rgb}{0.2,0.6,0.2}
%\definecolor{vermelinho}{rgb}{0.7,0.3,0.4}
%\definecolor{estranho}{cmyk}{0.4,1,0,0.4}
%\definecolor{brown}{cmyk}{0.25,0.35,0.75,0.3}
%\definecolor{marrom}{rgb}{0.7,0,0.8}
%%%%%%%%%%%%%%%%%%%%%%%%%%%%%%%%%%%%%%%%%%%%%%%%%%%%%%%%%%%%%%%%%%%%%%%%%%%%%%%%%%%%%
%\newcommand{\azul}{\color{blue}}
%\newcommand{\verde}{\color{green}}
%\newcommand{\vermelho}{\color{red}}
%\newcommand{\verm}{\color{red}}
%\newcommand{\roxo}{\color{estranho}}
%\newcommand{\amarelo}{\color{yellow}}
%\newcommand{\marrom}{\color{marrom}}
%%%%%%%%%%%%%%%%%%%%%%%%%%%%%%%%%%%%%%%%%%%%%%%%%%%%%%%%%%%%%%%%%%%%%%%%%%%%%%%%%%%%%
\newcommand{\be}{{\mathbf{e}}}
\newcommand{\bom}{{\mathbf{\omega}}}
\newcommand{\bT}{{\mathbf{T}}}
\newcommand{\bR}{{\mathbf{R}}}
\newcommand{\bA}{{\mathbf{A}}}
\newcommand{\bB}{{\mathbf{B}}}
\newcommand{\bD}{{\mathbf{D}}}
\newcommand{\bF}{{\mathbf{F}}}
\newcommand{\bd}{{\mathbf{d}}}
\newcommand{\bx}{{\mathbf{x}}}
\newcommand{\by}{{\mathbf{y}}}
%%%%%%%%%%%%%%%%%%%%%%%%%%%%%%%%%%%%
\newcommand{\at}{{\AA_\theta}}

%%%%%%%%%%%%%%%%%%%%%%%%%%%%%%%%%%%%%%%
%%%%%%%%%%%%%%%%%%%%%%%%%%%%%%%%%%%%%%%
%%%%%%%%%%%%%%%%%%%%%%%%%%%%%%%%%%%%%%%
%%%%%%%%%%%%%%%%%%%%%%%%%%%%%%%%%%%%%%%
%%%%%%%%%%%%%%%%%%%%%%%%%%%%%%%%%%%%%%%
%%%%%%%%%%%%%%%%%%%%%%%%%%%%%%%%%%%%%%%

\title{Complete loop quantization of a dimension 1+2 
Lorentzian gravity theory}

\author{Rodrigo M S Barbosa, Clisthenis P Constantinidis,
Zui Oporto and Olivier Piguet}

\address{Departamento de F\ii sica, Universidade Federal do Esp\ii rito Santo, 
Vit\'oria, ES, Brazil}

\ead{rodrigo\_martins@email.com, 
cpconstantinidis@gmail.com, 
azurnasirpal@gmail.com, opiguet@yahoo.com}

\begin{abstract}
De Sitter Chern-Simons gravity in $D=1+2$ spacetime is known to possess an extension 
with a Barbero-Immirzi like parameter. We find a partial gauge fixing 
which leaves a compact residual gauge group, namely SU(2). 
The compacticity of the residual gauge group opens the way to 
the usual LQG quantization techniques. We recall the exemple of 
the LQG quantization of SU(2) CS theory with cylindrical space 
topology, which thus provides a complete LQG of a 
Lorentzian gravity model in 3-dimensional space-time. 
\end{abstract}

%%%%%%%%%%%%%%%%%%%%%%%%%%%%%%%%%%%%%%%%%%%%%%%%%%%%%%%%%%%%%%%%%%%
%%%%%%%%%%%%%%%%%%%%%%%%%%%%%%%%%%%%%%%%%%%%%%%%%%%%%%%%%%%%%%%%%%%%
\section{Introduction}
The gauge group of $D$-dimensional Lorentzian spacetime, the Lorentz group SO(1,$D$-1),
is {\it noncompact}.  In the loop quantization 
framework\footnote{See~\cite{general-ref} for 
general references on the loop quantization of General Relativity.},
this is known to make difficult a  
proper definition of the internal
product in the vector space of quantum states.

For space-time dimension $1+3$, the usual way out is, starting from the Palatini-Holst 
action~\cite{holst},
 to partially fix the gauge (the {\it ``time
gauge''}) in such a way that the residual group be the {\it compact}
group SO(3) or SU(2). The presence of the {\it Barbero-Immirzi
parameter}~\cite{barbero-immirzi}
$\g$ appears  to be crucial. 
However, for other dimensions, with the Lorentz
group being SO($D$,1), the time gauge does not lead to the compact gauge group
SO($D$). 
But a recent proposal has been given
by the authors of Ref.~\cite{thiemann-D>4}, who have shown that it
exists a Hamiltonian framework where the gauge group may be chosen to
be compact, e.g., SO($D$), even in the case of a Lorentzian theory with
$(1,\,D-1)$ signature. In the latter case, however, as 
these authors have pointed out, their
construction is not possible in a  Lagrangian framework. The aim of
the present talk is to show that a compact gauge group  quantization
does exist in a Lorentzian space-time of dimension $1+2$ -- where also the time gauge 
dos not help --
provided there is a positive cosmological constant.

Our starting point is a formulation, due to Bonzom and
Livine~\cite{bonzom-livine}, of $(1+2)$-gravity with cosmological constant, in the 
presence of a Barbero-Immirzi like parameter. The gauge group of this theory is the de
Sitter group  SO(1,3),  and we shall show the existence of a partial gauge-fixing which reduces the
theory to a usual Chern-Simons theory with the {\it compact} SU(2) gauge
group. One 
can then apply known results in order to quantize the theory. One may recall, 
e.g., the LQG results of~\cite{Constantinidis-Luchini-Piguet}, obtained by
explicitly solving all the constraints, in a particular topology 
of 2-dimensional space.

Our case is somewhat different from that of the authors of~\cite{thiemann-D>4} (second paper). On one
hand, their simplicity constraints are not needed in $1+2$ dimensions;
on the other hand, we rely strongly on the existence of the Barbero-Immirzi
like  parameter of Bonzom and Livine, a feature very peculiar to that dimension.
Moreover, here we have -- and must have -- a non-zero cosmological constant.

%%%%%%%%%%%%%%%%%%%%%%%%%%%%%%%%%%%%%%%%%%%%%%%%%%%%%%%%%%%%%%%%%%%%
\section{(2+1)-Gravity with a cosmological constant as a Chern-Simons theory
with a Barbero-Immirzi parameter} 

The variables of 1+2 gravity in the first order formalism\footnote{The
tangent space
indices $I,J,\cdots= 0,1,2$ are lowered and raised with the Minkowski metric
$\eta_{IJ} = \mbox{diag}(-1,1,1)$ and with its inverse $\eta^{IJ}$. $\mu,\n,\cdots = t,x,y$
are world coordinates indices, and the space-time metric reads 
$g_{\m\n}=\eta_{IJ}e^I_\m e^J_\n$.}
 are the triad
forms $e^I=e^I_\m dx^\m$ and the Lorentz conections forms 
$\om_I=\half \varepsilon_{IJK}\,\om^{JK} = \om_{I\m} dx^\m$.

We suppose that we have a positive cosmological constant $\LA>0$. The gauge invariance group of 
the theory is then the de Sitter gauge group SO(1,3). Note that if $\LA$ were negative, the 
gauge group would be SO(2,2) and our procedure for obtaining a compact 
residual gauge group
would fail. The basis for the Lie algebra so(1,3) is given by the three
Lorentz generators $J^I\equiv\half\varepsilon^{IJK}J_{JK}$  and the three ``translation''
generators $P_I$, obeying the commutation rules\footnote{The Levi Civita
tensor $\varepsilon_{IJK}$ is normalized as $\varepsilon_{012}=1$.
Moreover, $\varepsilon^{IJ}{}_K=\eta^{IM}\eta^{JN}\varepsilon_{MNK}$,
etc. Beware of the signs!}
\[
[J^I,J^J] = \varepsilon^{IJ}{}_K J^K\,,\quad   [J^I,P_J] = \varepsilon^I{}_J{}^K P_K\,,\quad
[P_I,P_J] =  \LA\varepsilon_{IJK} J^K\,.
\]
We shall denote by $A=\om_I J^I + e^I P_I$ the SO(1,3) connection,
obeying the gauge transformation rules
$
A'(x)= g^{-1}(x)dg(x) + g^{-1}(x)A(x)g(x)\,, g\in\mbox{SO(1,3)}$
which, when written in terms of the component fields, reproduce 
the well-known de Sitter transformation rules.

Following~\cite{bonzom-livine}, we shall start from the fact that the most general 
background independent and gauge invariant action depending only on $A$
is of the Chern-Simons form and has two independent terms:
\eq
S = 
-\dfrac{\ka}{2}\dint_{\MM}\langle A,\,dA+\dfrac{2}{3}A\,A\rangle_1 
-\dfrac{\ka}{\g}\dint_{\MM}\langle A,\,dA+\dfrac{2}{3}A\,A\rangle_2 \,,
\eqn{action} 
($\wedge$ symbol omitted)
where $\vev{\cdot,\cdot}_\a\,,\a=1,2$ are the two invariant quadratic
forms corresponding to the two quadratic Casimir operators of
SO(1,3)~\cite{witten1}:
$C_{(1)} = P_IJ^I\,, 
C_{(2)}=\eta_{IJ}(\frac{1}{\LA}P^I P^J - J^I J^J)$.
%DAR O DETALHE DA A\CAO? (EST\'A NO ARQUIVO)
%\[\ba{l}
%S_{(1)} = -\dfrac{\ka}{2}\dint dt \dint_\S d^2x \lp \dot{\be}^I\bom_I
%+\dot\bom_I \be^I + 2e^I_t(\bR-\dfrac{\LA}{2} \be\times\be)_I + 2\om e^I_t\bT_I \rp
% \,,\esp
%S_{(2)} = -\dfrac{\ka}{2}\dint dt \dint_\S d^2x \lp -\LA \dot{\be}^I\be_I
%+\dot\bom_I \bom^I
%-2\LA e^I_t\bT_I + 2\om^I_t(\bR-\dfrac{\LA}{2} \be\times\be)_I\rp\,.
%\ea\]
%}
%{\bf Notation:} boldface type = forms in 2-space $\S$
%%\quad $a\cdot b = a^Ib_I= \eta_{IJ}a^Ib^I\,,quad 
%$(a\times b)_I = \varepsilon_I^{JK} a_Jb_K$\,,\\
%$\bR = \bd\bom + \bom\times\bom\,,\quad \bT = \bd\be+\bom\times\be$.
%     
%%%%%%%%%%%%%%%%%%%%%%%%%%%%%%%%%%%%%%%%%%%%%%%%%%%%%%%%%
The parameter $\ka$ is proportional to the inverse of the gravitation
constant, whereas
$\g$ is a parameter which shares with the usual Barbero-Immirzi parameter the
property of not appearing in the 
classical field equations, which read, indeed:
\[\ba{l}
F(A) = dA+A^2=0\,,\quad\mbox{or, in components:} \esp 
R^I\equiv
d\om^I+\varepsilon^I{}_{JK}\om^J \om^K = \dfrac{\LA}{2}\varepsilon^I{}_{JK}e^J e^K  \,,\quad 
T^I\equiv d e^I + \varepsilon^I{}_{JK}\om^Je^K =0\,.
\ea\]
One recognizes the Einstein equation with cosmological constant and the
null torsion condition.

%%%%%%%%%%%%%%%%%%%%%%%%%%%%%%%%%%%%%%%%%%%%%%%%%%%%%%%%%%%%%%%%%%%%
\section{Decomposition of SO(1,3) in ``rotations'' and ``boosts''}

We shall now introduce new variables corresponding to a decomposition
of SO(1,3) -- which is isomorphic to the 4-dimensional Lorentz group -- in 
``rotations'' and ``boosts''. We call $L_i$ the ``rotation'' generators
and ${ K_i}$  the ``boost'' generators ($i=1,2,3$).
\[
[L_i,L_j] = \varepsilon_{ijk} L_k\,,\quad [L_i,K_j] = \varepsilon_{ijk} K_k\,,\quad
[K_i,K_j] = - \varepsilon_{ijk} L_k\,.
\]
The new variables are the 1-forms $A^i$ and $B^i$ appearing in the
expression of the SO(1,3) connection: $\quad A= A^i L_i + B^i K_i$. One
recognizes in $A^i$ an SO(3) (or SU(2)) connection.

The relations between the old and new generators and variables read:
\[\ba{ll}
{ (L_1,\,L_2,\,L_3)} = (P_2/\sqrt{\LA},\, -P_1/\sqrt{\LA},\,-J^0)\,,\quad&
{ (K_1,\,K_2,\,K_3)} = (J^2,\, -J^1,\,P_0/\sqrt{\LA})\esp
{ (A^1,\,A^2,\,A^3)} = (\sqrt{\LA}e^2,\,-\sqrt{\LA}e^1,\, -\om_0) \,,\quad&
{ (B^1,\,B^2,\,B^3)} = (\om_2,\,-\om_1,\,\sqrt{\LA}e^0)
\ea\]
Our partial gauge fixing will consist in freezing the ``boost'' gauge degrees of freedom, 
keeping SU(2) as a residual gauge invariance. 

Let us first write the action
in the new variables\footnote{As usual we assume that space-time $\MM$ may
be split as $\real\times\S$, where $\S$ is the 2-dimensional space
manifold. Boldface letters mean forms, etc. in $\S$.}:
\[\ba{l}
S= -\dfrac{\ka}{2}\dint_\real dt \dint_\S \lp \dot\bB{}^i(\bA^i+\bB^i/\g)
+ \dot\bA{}^i(\bB^i-\bA^i/\g)\rp - \dint_\real dt\;H \,,\esp
\ea\]
where $H$ is the Hamiltonian $\ H=\int d^2x\lp A_t\GG_A(\bx) +
B_t\GG_B(\bx)\rp$.
The kinetic terms of the action determine the symplectic structure
of the theory, or equivalently its Poisson bracket algebra, the non-vanishing 
brackets being:
\[\ba{c}\{A^i_a(\bx),\,A^j_b(\by)\}
 = \dfrac{1}{\ka}\varepsilon_{ab}\d^{ij}\dfrac{\g}{\g^2+1} \d^2(\bx-\by)\,,
\quad \{B^i_a(\bx),\,B^j_b(\by)\}
 = -\dfrac{1}{\ka}\varepsilon_{ab}\d^{ij}\dfrac{\g}{\g^2+1} \d^2(\bx-\by)\,, \esp
\{B^i_a(\bx),\,A^j_b(\by)\}
 = -\dfrac{1}{\ka}\varepsilon_{ab}\d^{ij}\dfrac{\g^2}{\g^2+1} \d^2(\bx-\by)\,.
\ea\]
The Hamiltonian is purely constraints, the fields $A_t$ and $B_t$ 
playing the role of Lagrangian multipliers for
the Gauss and curvature constraints
\[\ba{ll}
\GG_A(\varepsilon) = \ka\dint_\S
\varepsilon^i\lp \bD\bB 
- \dfrac{1}{\g}\lp\bF(\bA)-\half\bB\times\bB\rp \rp^i\approx 0\,,
\quad\quad&\mbox{with}\ \ \bD=\bd+\bA\times\esp
\GG_B(\eta) = \ka\dint_\S
\eta^i\lp \bF(\bA)-\half\bB\times\bB+\dfrac{1}{\g}\bD\bB \rp^i \approx 0\,,
&\mbox{and}\ \ \bF(\bA)=\bd\bA+\half\bA\times\bA \,.
\ea\]
These constraints are first class:
\[\ba{l}
\{\GG_A(\varepsilon)\,\,\GG_A(\varepsilon')\} = \GG_A(\varepsilon\times\varepsilon')\,,
\     \{\GG_A(\varepsilon)\,\,\GG_B(\eta)\} = \GG_B(\varepsilon\times\eta)\,,
\     \{\GG_B(\eta)\,\,\GG_B(\eta')\} = - \GG_A(\eta\times\eta')\,,
\ea\]
and generate the gauge transformations:
\[%\ba{ll}
\{\GG_A(\varepsilon),\,\bA\} = \bD\varepsilon\,,\ \    \{\GG_B(\eta),\,\bA\} = 
   - \eta\times\bB\,,\ \
\{\GG_A(\varepsilon),\,\bB\} = \varepsilon\times\bB  \,, \ \ 
\{\GG_B(\eta),\,\bB\} =    \bD\eta  \,. 
%\ea
\]

%%%%%%%%%%%%%%%%%%%%%%%%%%%%%%%%%%%%%%%%%%%%%%%%%%%%%%%%%%%%%%%%%%%%
\section{Partial gauge fixing}
We choose to impose the axial-like gauge condition $B^i_y=0$,
i.e., $\om_{1y} = \om_{2y} = e^0_{y} = 0$,
 adding to the Hamiltonian 
a term $\ H\rightarrow H + \int_\S d^2x\,\m^i B^i_y$, the field $\m$ being a 
Lagrange multiplier. It turns out that the constraints $B^i_y \approx 0$ 
and $\GG_B\approx0$ are second class,
whereas $\GG_A\approx0$ remains first class. Following Dirac~\cite{dirac}, 
we introduce the corresponding Dirac
brackets, and treat the second class constraints as strong equalities.
In particular, $\GG_B=0$ yields $D_yB^i_x - \g  F^i_{xy}=0$, which
implies that $B^i_x$ is not an independent field.  

The set o independent dynamical variables  reduces to the pairs of conjugate variables
$A^i_x$ and $A^i_y$, whose Dirac brackets read
\[
\{A^i_x(\bx),\,A^j_y(\bx')\}_{\rm D} = \dfrac{1}{\ka}
\d^{ij}\dfrac{\g}{\g^2+1}\d^2(\bx-\bx')\,.
\]
With the field redefinition
$
\AA^i_x=A^i_x-\g B^i_x\,,\quad \AA^i_y=A^i_y\,,
$
the brackets and the hamiltonian read
\eq
\{\AA^i_x(\bx),\,\AA^j_y(\bx')\}_{\rm D} = \dfrac{\g}{\ka}
\d^{ij}\d^2(\bx-\bx')
\quad\mbox{and}\quad
H = -\dfrac{\ka}{\g} \dint_\S d^2x\, A^i_t F^i_{xy}(\AA)\,.
\eqn{2}
These are  the Hamiltonian and brackets of a Chern-Simons theory 
for the {\it compact} gauge group SU(2).

%%%%%%%%%%%%%%%%%%%%%%%%%%%%%%%%%%%%%%%%%%%%%%%%%%%%%%%%%%%%%%%%%%%%
\section{Quantization}

The problem of the quantization of the present theory is thus reduced to
that of the Chern-Simons theory with gauge group SU(2). We may
refer for this to the literature~\cite{witten1,witten2,quantiz-of-CS}. The residual gauge 
symmetry group SU(2) being compact, LQG methods apply. The case of 2-dimensional
space being a cylinder was treated in~\cite{Constantinidis-Luchini-Piguet}, 
with a complete solution of the constraints.

%%%%%%%%%%%%%%%%%%%%%%%%%%%%%%%%%%%%%%%%%%%%%%%%%%%%%%%%%%%%%%%%%%%%
\section{Conclusions}

We have thus succeeded to reduce the gauge symmetry to that of a compact
group, namely SU(2), through a suitable gauge fixing. We note that
the presence of a Holst-like term in the action (second term in 
\equ{action}), together with the Barbero-Immirzi-like parameter $\g$, is crucial,
much in the same way as in the $D=3+1$ case, where the gauge symmetry
of the Palatini-Holst action~\cite{holst} is reduced to the same SU(2) through a time
gauge fixing -- which is only available in that dimension!

A complete quantization of 3 dimensional gravity with a positive cosmological constant
can thus be performed in the usual loop quantization scheme, thanks 
to the compactness of SU(2).

A more detailed account will appear in~\cite{rodrigo-clist-zui-op}.

%%%%%%%%%%%%%%%%%%%%%%%%%%%%%%%%%%%%%%%%%%%%%%%%%%%%%%%%%%%%%%%%%%%%

%%%%%%%%%%%%%%%%%%%%%%%%%%%%%%%%%%%%%%%%%%%%%%%%%%%%%%%%

\section*{References}
\begin{thebibliography}{99}

\bibitem{general-ref} C. Rovelli, ``Quantum Gravity'', Cambridge Monography on Math.
	Physics (2004);
A. Ashtekar and J. Lewandowski, ``Background independent quantum gravity: 
        A status report'', \cqg{21}{2004}{R53}) [arXiv:gr-qc/0404018];
T. Thiemann,
 	``Modern Canonical Quantum General Relativity'',
   Cambridge Monographs on Mathematical Physics (2008);
M. Han, W. Huang and Y. Ma ``Fundamental structure of loop quantum gravity'',
   \ijmp{D16}{2007}{1397}, [arXiv:gr-qc/0509064]; 
H. Nicolai, K. Peeters and M. Zamaklar, 
   ``Loop quantum gravity: An outside view'', 
   \cqg{22}{2005}{R193},
	 [arXiv:hep-th/0501114]; 
H. Nicolai and K. Peeters, ``Loop and spin foam quantum gravity: A brief
   guide for beginners'', 	 [arXiv:gr-qc/0601129]; 
T. Thiemann, 	``Loop quantum gravity: An inside view'',
	 [arXiv:hep-th/0608210].

\bibitem{holst} S. Holst, ``Barbero's Hamiltonian derived from a generalized 
Hilbert-Palatini action'', Phys.Rev. D53 (1996) 5966-5969, 
[arXiv: gr-qc/9511026].

\bibitem{barbero-immirzi} J.F. Barbero,  ``Reality conditions and Ashtekar
variables: A Different perspective'', Phys. Rev. D51 (1995) 5507, 
[arXiv: gr-qc/9410013]; 
     Giorgio Immirzi, ``Real and complex connections for canonical
gravity'',  Class.Quant.Grav. 14 (1997) L177-L181, 
[arXiv: gr-qc/9612030]. 

\bibitem{thiemann-D>4} N. Bodendorfer, T. Thiemann and A. Thurn
``New Variables for Classical and Quantum Gravity in all Dimensions
I,II,III and IV'', 
[arXiv:1105.3703 - 1105.3706 [gr-qc]]; 
N. Bodendorfer, in these proceeedings; 
T. Thiemann, in these proceeedings; 
A, Thurn, in these proceeedings.

\bibitem{bonzom-livine} V. Bonzom and E.R. Livine, 
``A Immirzi-like parameter for 3d quantum gravity''
\cqg{25}{2008}{195024}, [arXiv:0801.4241[gr-qc]]

\bibitem{Constantinidis-Luchini-Piguet} C.P. Constantinidis, G. Luchini and O. Piguet,
``The Hilbert space of Chern-Simons theory on the cylinder.
A Loop Quantum Gravity approach'', \cqg{27}{2010}{065009},
[arXiv:0907.3240[gr-qc]].

\bibitem{witten1} E. Witten; ``$2+1$ dimensional Gravity as an Exactly Soluble system'', 
\np{B311}{1988}{46}. 

\bibitem{dirac}   P.A.M. Dirac, ``Lectures on Quantum Mechanics'',
Dover, 2001;   
M. Henneaux, C. Teitelboim, ``Quantization of Gauge Systems'',
Princeton University Press, 1994.

\bibitem{witten2} 
E.Witten ``Quantum Field Theory and the Jones Polynomial´´, \cmp{121}{1989}{351}. 

\bibitem{quantiz-of-CS} G.V. Dunne, R. Jackiw, C.A. Trugenberger,
  ``Chern-Simons Theory in the Schr\"odinger Representation'',
  \annp{194}{1989}{197}; 
E.Guadagnini, M.Martellini, M.Mintchev, ``Braids and Quantum Group Symmetry in 
Chern-Simons Theory´´, \np{B336}{1990}{581};
Steven Carlip, ``Quantum Gravity in 2+1 Dimensions'', 
Cambridge Monographs on Mathematical Physics (2003).

\bibitem{rodrigo-clist-zui-op} R.M.S. Barbosa, C.P. Constandinis, Z. Oporto and O. Piguet,
in preparation.


\end{thebibliography}
\end{document}